\documentclass[12pt,letter,draftcls,a4paper,onecolumn]{IEEEtran}

\usepackage{dsfont}
\usepackage{amsmath}
\usepackage{amsthm}
\usepackage{amssymb}
\usepackage{amsfonts}
\usepackage{graphicx}
\usepackage{subfigure}
\usepackage{color}
\usepackage{amsmath}
\usepackage[all,poly]{xy}
\usepackage{multirow}
\usepackage{algorithm}
\usepackage{algorithmic}
\usepackage[numbers, square, comma, sort&compress]{natbib}
\usepackage{epic,eepic,pstricks}
\usepackage{bm}
\usepackage{lineno}

\graphicspath{{figures/}}

\newcommand{\btheta}{\boldsymbol{\theta}}
\newcommand{\bz}{\boldsymbol{z}}

\theoremstyle{plain}

\title{Computing the Cramer-Rao bound of Markov random field parameters: Application to the Ising and the Potts models}
\author{Marcelo Pereyra, Nicolas Dobigeon, Hadj Batatia and Jean-Yves
Tourneret
\thanks{Marcelo Pereyra gratefully acknowledges financial support from the SuSTaIN program - EPSRC grant EP/D063485/1 - at the Department of Mathematics, University of Bristol, and from a postdoctoral fellowship from French Ministry of Defence.
}
\thanks{Marcelo Pereyra is with Department of Mathematics of the University of Bristol, UK
e-mail: marcelo.pereyra@bristol.ac.uk}
\thanks{Nicolas Dobigeon, Hadj Batatia and Jean-Yves
Tourneret are with University of Toulouse, IRIT/INP-ENSEEIHT/T\'eSA,
2 rue Charles Camichel, BP 7122, 31071 Toulouse cedex 7, France
e-mail:\{Nicolas.Dobigeon,Hadj.Batatia,Jean-Yves.Tourneret\}@enseeiht.fr.}
}

\begin{document}

\maketitle

\begin{abstract}
This letter considers the problem of computing the Cramer-Rao bound for the parameters of a Markov random field. Computation of the exact bound is not feasible for most fields of interest because their likelihoods are intractable and have intractable derivatives. We show here how it is possible to formulate the computation of the bound as a statistical inference problem that can be solve approximately, but with arbitrarily high accuracy, by using a Monte Carlo method. The proposed methodology is successfully applied on the Ising and the Potts models.
\end{abstract}
\begin{IEEEkeywords}
Cramer Rao bound, Monte Carlo algorithms, Markov random fields, Intractable distributions.
\end{IEEEkeywords}

\section{Introduction}
\label{sec:intro} The estimation of parameters involved in intractable statistical models (i.e., with intractable likelihoods) is a difficult problem that has received significant attention in the recent computational statistics and signal processing literature \cite{Moller2006,Murray2006,Andrieu2009,Andrieu2010}. Particularly, estimating the parameters of a Markov random field (MRF) is an active research topic in image processing \cite{Descombes1999,Forbes2007,McGrory2009,Giovannelli,Pereyra_TIP_2012}. Several new unbiased estimators have been recently derived, mainly based on efficient Monte Carlo (MC) methods \cite{Moller2006,Murray2006,Pereyra_TIP_2012,Everitt2012}.

This letter addresses the problem of computing the Cramer-Rao bound (CRB) \cite{VanTreesBook} for estimators of MRF parameters. Knowing the CRB of a statistical model is of great importance for both theoretical and practical reasons. From a theoretical point of view, the CRB establishes a lower limit on how much information a set of observations carries about unknown parameters. Specifically, it defines the minimum variance for any unbiased estimator of these parameters. From a practical perspective, the CRB is used as a means to characterize the performance of unbiased estimators in terms of mean square error (i.e., estimation variance). Unfortunately, the CRB for most MRF models is difficult to compute because their likelihoods are intractable \cite[Chap. 7]{LiBook}. 

This letter addresses this difficulty by formulating the computation of the CRB as a statistical inference problem that can be solved using MC methods \cite{Robert}. Precisely, we propose to express the CRB in terms of expectations that can be efficiently estimated by MC integration \cite[Chap. 3]{Robert}. The proposed CRB estimation method is demonstrated on specific MRF models that have been widely used in the image processing community, namely the Potts and Ising models. The remainder of the letter is organized as follows: Section II defines the class of statistical models considered in this work and proposes an original method to estimate their CRB based on a MC algorithm. The application of the proposed methodology to the Ising and Potts models is presented in Section III. Conclusions are finally reported in Section IV.

\section{Computing the Fisher Information Matrix $\mathcal{I}(\boldsymbol{\theta})$}\label{Computing}
Let $\boldsymbol{\theta} = \left(\theta_1,\ldots,\theta_M\right)^T$ be
an unknown parameter vector and $\boldsymbol{z} =
\left(z_1,z_2,\ldots,z_N\right)^T$ an observation vector whose
elements take their values in a set $\Omega$. This paper considers
the case where $\boldsymbol{\theta}$ and $\boldsymbol{z}$ are
related by the following generic distribution
\begin{equation} \label{Gibbs}
f_{\btheta}(\bz) \triangleq
\frac{1}{C(\boldsymbol{\theta})}
\exp{\left[\boldsymbol{\theta}^T\Phi(\boldsymbol{z})\right]}
\end{equation}
where $\Phi (\boldsymbol{z}) : \Omega^N \rightarrow \mathbb{R}^M$
is a sufficient statistic and $C(\boldsymbol{\theta})$ is the
normalizing constant given by
\begin{equation} \label{C}
C(\boldsymbol{\theta}) =
\int_{\Omega^N}\exp{\left[\boldsymbol{\theta}^T\Phi(\boldsymbol{z})\right]}d\boldsymbol{z}
\end{equation}
where integration is performed in the Lebesgue sense with respect to an appropriate measure on $\Omega^N$. Note that the model \eqref{Gibbs} defines an important subclass of the exponential family. It comprises several standard distributions, such as Gaussian, Laplace or gamma distributions, as well as multivariate distributions frequently used in signal and image processing applications, such as Markov random fields \cite{LiBook}. In this latter case, the normalizing constant, (also known as partition fuction), is generally intractable owing to the inherent difficulty of evaluating integrals over $\Omega^N$ when $N$ is large \cite{LiBook}.

The Cramer-Rao bound of $\boldsymbol{\theta}$ establishes a lower
bound on the covariance matrix of any unbiased estimator
$\hat{\boldsymbol{\theta}}$ of $\boldsymbol{\theta}$ \cite{VanTreesBook}. Because the
existence of the bound requires that
$f_{\boldsymbol{\theta}}(\boldsymbol{z})$ verifies some weak
regularity conditions, we will assume that $C(\boldsymbol{\theta})$
is continuously differentiable (i.e., $C(\boldsymbol{\theta})
\in \mathcal{C}^1$). Then the CRB is equal to the inverse of the Fisher information matrix (FIM) of $\boldsymbol{\theta}$ \cite{VanTreesBook}, i.e.,
\begin{equation}\label{CRB}
\textrm{cov}(\hat{\boldsymbol{\theta}}) \geq CRB(\btheta) =
\left[\mathcal{I}(\boldsymbol{\theta})\right]^{-1}
\end{equation}
where the inequality \eqref{CRB} means that the matrix $\textrm{cov}(\hat{\boldsymbol{\theta}}) - \left[\mathcal{I}(\boldsymbol{\theta})\right]^{-1}$ is positive semidefinite and where $\mathcal{I}(\boldsymbol{\theta})$ is an $M \times M$ positive semidefinite symmetric matrix whose element $(i,j)$ is given by \cite{VanTreesBook}
\begin{equation}\label{FIM}
\mathcal{I}_{i,j}(\boldsymbol{\theta}) \triangleq \mathrm{E}_{f}\left[\frac{\partial}{\partial \theta_i}
\log[f_{\boldsymbol{\theta}}(\boldsymbol{z})] \frac{\partial}{\partial \theta_j}
\log[f_{\boldsymbol{\theta}}(\boldsymbol{z})]\right]
\end{equation}
where $\mathrm{E}_{f}$ denotes the expectation operator with respect to $f_{\boldsymbol{\theta}}(\boldsymbol{z})$. By applying the definition \eqref{FIM}
to \eqref{Gibbs} we obtain that
\begin{equation}\label{FIM2}
\begin{split}
\mathcal{I}_{i,j}(\boldsymbol{\theta}) = \mathrm{E}_{f}\left[\left(\Phi_i(\boldsymbol{z}) - \frac{\partial}{\partial \theta_i}
\log C(\boldsymbol{\theta})\right)\left(\Phi_j(\boldsymbol{z})- \frac{\partial}{\partial \theta_j}
\log C(\boldsymbol{\theta})\right)\right]
\end{split}
\end{equation}
where $\Phi_i(\boldsymbol{z}) : \Omega^N \rightarrow \mathbb{R}$ is the $i$th component of the vector field $\Phi(\boldsymbol{z}) = \left[\Phi_1(\boldsymbol{z}),\ldots,\Phi_M(\boldsymbol{z})\right]^T$. Unfortunately, evaluating \eqref{FIM2} for MRF models is rarely possible because of the intractability of the derivatives $\frac{\partial}{\partial \theta_i}\log C(\boldsymbol{\theta})$. Note that difficulty cannot be addressed by numerical differentiation because $\log C(\boldsymbol{\theta})$ is itself intractable.

This letter proposes to exploit a property of the exponential family to replace the intractable derivatives in $\mathcal{I}(\boldsymbol{\theta})$ by expectations that can be efficiently approximated using MC integration \cite[Chap. 3]{Robert}. Precisely, we use the following property that relates the derivatives $\frac{\partial}{\partial \theta_i}\log C(\boldsymbol{\theta})$ to the expectations of $\boldsymbol{\Phi}$ \cite[p. 118]{BayesianChoice}
\begin{equation}\label{identity}
\begin{split}
\frac{\partial}{\partial \theta_i} \log[C(\boldsymbol{\theta})]
&= \frac{1}{C(\boldsymbol{\theta})} \int_{\mathcal{S}^N}\Phi_i(\boldsymbol{z})\exp{\left[\boldsymbol{\theta}\Phi(\boldsymbol{z})\right]}d\boldsymbol{z}\\
&= \mathrm{E}_{f}\left[\Phi_i(\boldsymbol{z})\right].
\end{split}
\end{equation}
Again, integration is performed in the Lebesgue sense with respect to an appropriate measure on $\Omega^N$. By substituting property \eqref{identity} in equation \eqref{FIM2} we obtain the matrix
\begin{equation}\label{FIMCov}
\mathcal{I}(\boldsymbol{\theta}) = \mathrm{cov}\left[\Phi(\boldsymbol{z})\right]
\end{equation}
whose elements
\begin{equation}\label{FIM4}
\mathcal{I}_{i,j}(\boldsymbol{\theta}) = \mathrm{E}_{f}\left[\Phi_i(\boldsymbol{z})\Phi_j(\boldsymbol{z})\right] -
\mathrm{E}_{f}\left[\Phi_i(\boldsymbol{z})\right]\mathrm{E}_{f}\left[\Phi_j(\boldsymbol{z})\right].
\end{equation}
The property \eqref{identity} has been used previously to derive the statistical moments of $\boldsymbol{\Phi}$ from the derivatives of $C(\btheta)$ in cases where $C(\btheta)$ is known and can be differentiated analytically \cite[p. 116]{Bishop} as well as to study the physical properties of lattice systems \cite[Chap. 31]{MacKay}. However, to the best of our knowledge this is the first time that this property is used in a statistical inference context to compute a CRB.

Expression \eqref{FIM4} differs from \eqref{FIM2} by the fact that derivatives are replaced by expectations of $\boldsymbol{\Phi}$ w.r.t. $f_{\btheta}(\bz)$. From a computational perspective this alternative expression is fundamentally better than \eqref{FIM2} because the expectations, in spite of being intractable, can be efficiently approximated with arbitrarily high accuracy by MC integration \cite[Chap. 3]{Robert}. Note that MC approximations are particularly well suited for high-dimensional models given that their accuracy depends exclusively on the number of MC samples used and not on the dimension of the model.

Lastly, approximating $\mathcal{I}(\boldsymbol{\theta})$ by MC integration requires simulating samples distributed according to $f_{\btheta}(\bz)$. In this letter, this is achieved by using a Gibbs sampler that admits $f_{\btheta}(\bz)$ as unique stationary distribution \cite{Robert}. This sampler belongs to the class of Markov chain Monte Carlo algorithms, which are interesting for MRFs because they not require to know $C(\boldsymbol{\theta})$. The output of this algorithm is a Markov chain of $N_{MC}$ samples ${\{\bz^{(t)}\}}_{t=1}^{N_{MC}}$ that can be used to approximate $\mathcal{I}(\boldsymbol{\theta})$ through the sample covariance matrix
\begin{equation}\label{FIMhat}
\hat{\mathcal{I}}(\boldsymbol{\theta}) = \frac{1}{N_{MC}-1} \sum_{t=1}^{N_{MC}} \left(\boldsymbol{u}^{(t)} - \bar{\boldsymbol{u}}\right)\left(\boldsymbol{u}^{(t)} - \bar{\boldsymbol{u}}\right)^T, \quad \boldsymbol{u}^{(t)} = \Phi(\boldsymbol{z}^{(t)})
\end{equation}
with $\bar{\boldsymbol{u}} = \sum_{t=1}^{N_{MC}} \boldsymbol{u}^{(t)}/N_{MC}$. The proposed MC algorithm is summarized in Algo. \ref{algo:MonteCarlo} below.

\begin{algorithm}
\caption{MC algorithm} \label{algo:MonteCarlo}
    \begin{algorithmic}[1]
    \STATE Input: $\boldsymbol{\theta}$, initial condition $\boldsymbol{z}^{(0)}$, number of Monte Carlo samples $N_{MC}$.
    \FOR{$t = 1$ to $N_{MC}$}
    \STATE Generate $\boldsymbol{z}^{(t)} \sim K_{\boldsymbol{\theta}}(\boldsymbol{z}^{(t-1)}|\cdot)$
    \STATE Set $\boldsymbol{u}^{(t)} = \Phi(\boldsymbol{z}^{(t)})$
    \ENDFOR
    \STATE Evaluate $\hat{\mathcal{I}}(\boldsymbol{\theta})$ using \eqref{FIMhat}
    \STATE Output $\hat{\mathcal{I}}(\boldsymbol{\theta})$
\end{algorithmic}
\end{algorithm}

The ergodicity of the Gibbs sampler guarantees that as the number of samples increases $\hat{\mathcal{I}}(\boldsymbol{\theta})$ converges to $I(\boldsymbol{\theta})$ (for details about MCMC algorithms and their practical application please see \cite{Robert,Geyer92}). Moreover, note that $\hat{\mathcal{I}}(\boldsymbol{\theta})$ is semipositive definite by construction and it is invertible whenever $N_{MC}$ is large enough such that $\{\boldsymbol{u}^{(t)}\}_{t=1}^{N_{MC}}$ spans $\mathbb{R}^M$. In practice this condition is satisfied almost surely if $N_{MC}$ is large enough to produce a stable estimate of $\mathcal{I}(\boldsymbol{\theta}$).

Finally, note that \eqref{FIMhat} is valid regardless of the specific MCMC method used to simulate from $f_{\btheta}(\bz)$. For generality this letter considers that samples are generated using a Gibbs sampler, which provides a general solution that it is easy to apply to any given MRF (Gibbs samplers are defined using the same distributions that are used to specify $f_{\btheta}(\bz)$, i.e., the conditional distributions of each element of $\boldsymbol{\theta}$ given the other elements $f_{\boldsymbol{\theta}}(z_i | z_1,\ldots,z_{i-1},z_{i+1},\ldots,z_N)$ \cite{LiBook}). However, the Gibbs sampler is not always the most efficient MCMC method to simulate from a specific MRF (e.g., the Ising and Potts models are more efficiently sampled with a Swendsen-Wang algorithm \cite{Swendsen87}).

\section{Application to the Ising and Potts Markov Random Fields}
\subsection{Ising and Potts models}
This section applied the proposed methodology to the computation of the CRB of two important intractable models, namely the homogeneous Ising and Potts MRF. For completeness these models are recalled below.

Let $\boldsymbol{z} = (z_1,z_2,\ldots,z_N)$ be a discrete random vector whose elements take their values in the finite set $\Omega_K
= \{1,\ldots,K\}$. The Ising and the Potts MRF are defined by the following probability mass function
\begin{equation} \label{Potts}
f_{\theta}(\boldsymbol{z}) \triangleq \frac{1}{C(\theta)}
\exp{\left[\theta\Phi(\boldsymbol{z})\right]}
\end{equation}
with
\begin{equation} \label{eq:potential}
    \Phi (\boldsymbol{z}) = \sum_{n=1}^N\sum_{n'\in \mathcal{V}(n)} \delta (z_{n} - z_{n'})
\end{equation}
where $\mathcal{V}(n)$ is the index set of the neighbors associated with the $n$th element, $\delta(\cdot)$ is the Kronecker function and $\theta
\in \mathbb{R}^+$ is the granularity coefficient or \emph{inverse temperature} parameter. The Gibbs distribution \eqref{Potts} corresponds to the Ising MRF when $K=2$, and to the Potts MRF for $K \geq 3$. In our experiments $\mathcal{V}(n)$ will be considered to be a bidimensional first-order (i.e., $4$-pixel) neighborhood structure. However, the proposed method is valid for any correct neighborhood structure (see \cite{LiBook} for more details). Finally, note that despite their simplicity these models are extensively used in modern image segmentation and/or classification applications (see \cite{Vincent2010,Eches2010tgrs,PereyraTMIC2011} and references therein) and that the estimation of the granularity parameter $\theta$ is still an active research topic \cite{Pereyra_TIP_2012}.

\subsection{Validation with ground truth}
To validate the proposed MC method under controlled conditions (i.e., for a known CRB), the proposed methodology has been first applied to an Ising model defined on a \emph{toroidal} graph (i.e., with cyclic boundary conditions) of size $N = 32 \times 32$. Unlike most MRF models, this particular MRF has a known normalizing constant and FIM \cite{Giovannelli}.

\begin{figure}[h!]
  \centerline{\includegraphics[width=12.0cm]{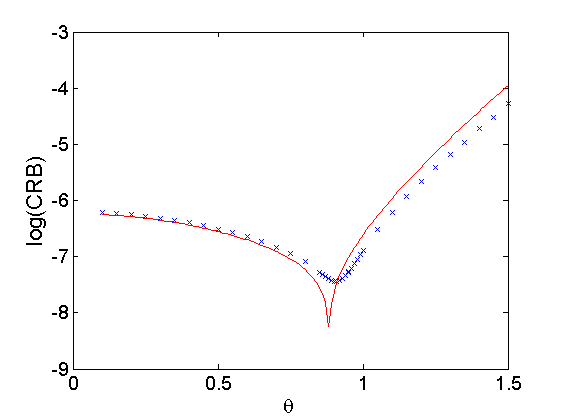}}
  \caption{Cramer-Rao bound for an Ising ($K=2$) defined on a toroidal graph. True CRB (solid red), estimates obtained by MC integration (blue crosses).}\label{fig:toroidal}
\end{figure}

Fig. \ref{fig:toroidal} compares the estimates obtained for different values of $\theta$ with the true CRB \cite{Giovannelli}. These estimates have been computed from Markov chains of $1\,000\,000$ samples generated with a Gibbs sampler (computing each estimate required $39$ minutes on a 2.6GHz Intel i7 quad-core workstation running MATLAB R2013a). To ease visual interpretation, results are displayed using a logarithmic scale.
We observe in Fig. \ref{fig:toroidal} that the estimates obtained with the proposed method are in good agreement with the true values of the CRB. We also observe that the error introduced by using a Monte Carlo approximation varies slightly with the value of $\theta$ and is larger at approximately $\theta = 0.9$, coinciding with the phase-transition temperature of the Ising MRF ($\theta_c = \log(1+\sqrt{2}) \approx 0.88$). These variations with $\theta$ are due to the fact that the mixing properties of the Gibbs sampler deteriorate at temperatures close to $\theta_c$ due to long range dependencies between the elements of $\bz$ \cite[p. 339]{Robert}. As a result the Markov chains associated with different values of $\theta$ have different effective samples sizes \cite[p. 499]{Robert} (i.e., different numbers of equivalent independent samples) and produce estimates with different accuracies. Indeed, the effective sample size, measured from the chain's autocorrelation function, is $870\,000$ samples for $\theta = 0.1$, it decreases progressively to $20\,000$ samples for $\theta = 0.9$ and then increases to $235\,000$ samples for $\theta = 1.5$.

\subsection{Asymptotic study of the CRB}
\begin{figure}[h!]
  \centerline{\includegraphics[width=12.0cm]{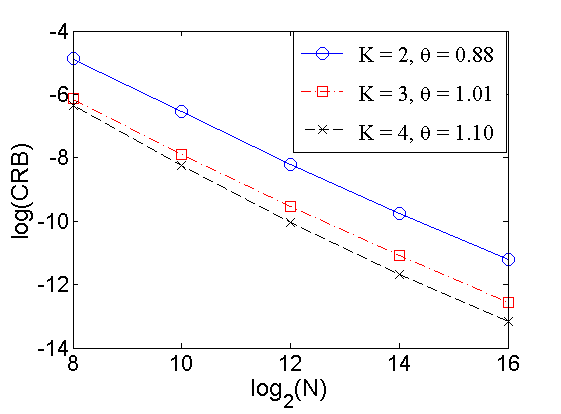}}
  \caption{Cramer-Rao bound for an Ising ($K=2$) and two Potts MRF ($K=3$ and $K=4$) close to phase-transition and for different field sizes $N$. Results are displayed in \emph{log-log} scales.}\label{fig:CRB}
\end{figure}

The second set of experiments shows the evolution of the CRB with respect to the size of observation vector $\boldsymbol{z}$ (i.e., the number of field components $N$). The CRB has been computed for the following $5$ field sizes $N = (2^8, 2^{10}, 2^{12}, 2^{14}, 2^{16})$, corresponding to bidimensional MRFs of size $16 \times 16$, $32 \times 32$, $64 \times 64$, $128 \times 128$ and $256 \times 256$. Experiments have been performed using an Ising MRF, a 3-state and a 4-state Potts MRF (i.e., $K=2$, $K=3$ and $K=4$ respectively) defined on a regular lattice (not a toroid). CRB estimates have been computed from Markov chains of $1\,000\,000$ samples, whose generation for the $32 \times 32$ and $K = 2,3,4$ cases required $39$, $43$ and $49$ minutes respectively on a 2.6GHz Intel i7 quad-core workstation running MATLAB R2013a. Finally, for each model, the parameter $\theta$ was set close to the critical phase-transition value, i.e., $\theta_c = \log(1 + \sqrt{K})$ to introduce a strong dependency between the components of the MRF. Fig. \ref{fig:CRB} shows the resulting CRBs versus the size of the MRF $N$ in logarithmic scales.

We observe that for all models the logarithm of the CRB decreases almost linearly with the logarithm of the number of field components. This result shows that the strong dependency between the field components does not modify significantly the linear behavior that is generally observed for models defined by statistically independent components. We also observe that the CRB decreases with the number of states $K$, indicating that an accurate estimation of $\theta$ for the Ising model is more difficult than for a Potts MRF.

\subsection{Evaluation of state-of-the art estimators of $\theta$}
The third set of experiments compares the CRB to the empirical variance of three state-of-the art estimation methods, the \emph{auxiliary variable} \cite{Moller2006}, \emph{exchange} \cite{Murray2006} and \emph{ABC} \cite{Pereyra_TIP_2012} algorithms. As explained previously, the CRB is often used as a means to measure the performance of unbiased estimators in terms of mean square error. In this letter, the three algorithms studied in \cite{Moller2006,Murray2006,Pereyra_TIP_2012} have been used to compute an approximate maximum-likelihood (ML) estimation of $\theta$ for the Ising and the $3$-state Potts MRF.

Experiments were conducted as follows. First $N_{ML} = 2\,500$ synthetic observation vectors $\boldsymbol{z}^{(i)} \sim f_\theta(\boldsymbol{z})$, $i = 1,\ldots,N_{ML}$ were generated using an appropriate Gibbs sampler. Then, for each observation $\boldsymbol{z}^{(i)}$, three ML estimates $\hat{\theta}_{EXCH}^{(i)}$, $\hat{\theta}_{ABC}^{(i)}$ and $\hat{\theta}_{AUX}^{(i)}$ were computed using the three estimation methods mentioned above. Precisely, each estimation method was used to generate a $1\,000$-sample MC approximation of the intractable posterior distribution $\pi(\theta|\boldsymbol{z}) = f_{\theta}(\boldsymbol{z}) \boldsymbol{1}_{[\theta_{\textrm{min}},\theta_{\textrm{max}}]}(\theta)/\pi(\boldsymbol{z})$, which is proportional to the likelihood $L(\theta | \bz) = f_{\theta}(\boldsymbol{z})$ and has the same maximizer within the range $[\theta_{\textrm{min}},\theta_{\textrm{max}}]$ (the values $\theta_{\textrm{min}} = 0$ and $\theta_{\textrm{max}} = \theta_{c} = \log(1+\sqrt{K})$ were used in the experiments). An ML estimate was computed by maximizing this MC approximation (we used Gaussian kernel smoothing to regularize the approximation). Finally, the variance of each estimator was approximated by computing the sample variance, e.g., $\textrm{Var}(\hat{\theta}_{ABC}) = (N_{ML}-1)^{-1} \sum_{i=1}^{N_{ML}}(\hat{\theta}_{ABC}^{(i)} - \bar{\theta}_{ABC})^2$ with $\bar{\theta}_{ABC} = N_{ML}^{-1}\sum_{i=1}^{N_{ML}} \hat{\theta}_{ABC}^{(i)}$. All algorithms used $250$ burn-in steps and $10$ Gibbs moves per auxiliary variable coordinate, which are realistic implementation conditions for signal processing applications \cite{Pereyra_TIP_2012}. Moreover, the auxiliary variable method \cite{Moller2006} was implemented using the true value of $\theta$ as auxiliary estimate, while the tolerance of the ABC method \cite{Pereyra_TIP_2012} was set to $1\%$ (see \cite{Pereyra_TIP_TechReport_2012} for more details about these methods and their application to the Ising and Potts MRFs).

\begin{figure}[h!]
\begin{minipage}[a1]{.99\linewidth}
  \centering
  \centerline{\includegraphics[width=12.0cm]{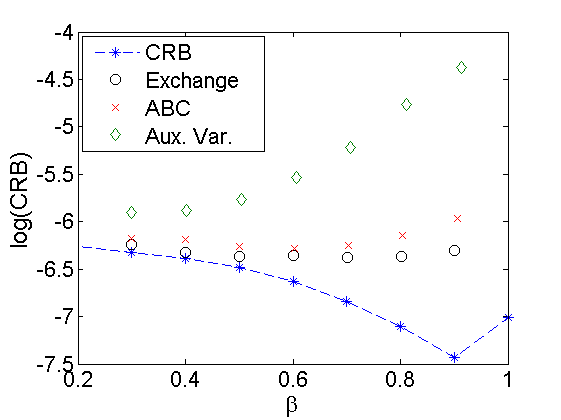}}
  \centerline{\small{(a) Ising MRF}}\medskip
\end{minipage}
\begin{minipage}[a1]{.99\linewidth}
  \centering
  \centerline{\includegraphics[width=12.0cm]{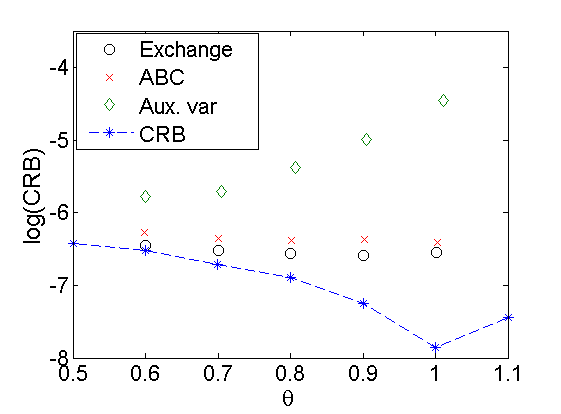}}
  \centerline{\small{3-state Potts MRF}}\medskip
\end{minipage}
\caption{Cramer-Rao bounds for an Ising and a $3$-state Potts models of size $32 \times 32$. Results are displayed in logarithmic scale.}\label{fig:Ising}
\end{figure}

Fig. \ref{fig:Ising}(a) compares the CRB estimated with our method for a Ising MRF of size $32 \times 32$ with the variance of the ML estimates obtained with the state-of-the art algorithms (the CRB estimates were computed from Markov chains of $1\,000\,000$ samples whose generation required $39$ minutes on a 2.6GHz Intel i7 quad-core workstation running MATLAB R2013a). These values have been computed for $\theta < \theta_c = \log(1+\sqrt{2})$ which is the range of interest for this model (for $\theta > \theta_c$ all the field components have almost surely the same color). We observe the good performance of the ML estimators based on the exchange \cite{Murray2006} and the ABC \cite{Pereyra_TIP_2012} for small values of $\theta$ (i.e., $\theta < 0.6$). However, their performance decreases progressively for $\theta > 0.6$, a behavior that is explained by the fact that the estimators use a Gibbs sampler to approximate the intractable likelihood. As explained previously, the mixing properties of this sampler deteriorate as $\theta$ increases towards the critical value $\theta_c$. This results in a degradation of the approximation of the likelihood and in a larger ML variance. Moreover, one can also see that the ML estimator based on the auxiliary variable method \cite{Moller2006} has a larger variance than the other estimators for all values of $\theta$. This result is in accordance with the experiments reported in \cite{Murray2006,Pereyra_TIP_2012}. Furthermore, Fig. \ref{fig:Ising}(b) shows the CRB computed for a $3$-state Potts MRF of size $32 \times 32$ and $\theta < \theta_c = \log(1+\sqrt{3})$ (these CRB estimates were computed from Markov chains of $1\,000\,000$ samples whose generation required $43$ minutes). Again, the CRB is compared to the variance of the ML estimates obtained with the state-of-the art algorithms. Similarly to Fig. \ref{fig:Ising}(a), the ML estimates based on the exchange \cite{Murray2006} and the ABC \cite{Pereyra_TIP_2012} methods are close to the CRB for small values of $\theta$ (i.e., $\theta < 0.8$), and depart progressively as $\theta$ approaches $\theta_c$ due to a degradation of the approximation of the likelihood.

\section{Conclusion}
\label{sec:conclusion} This letter studied the problem of computing the CRB for the parameters of Markov random fields. For these distributions the CRB depends on the derivatives of the normalizing constant or partition function $C(\boldsymbol{\theta})$, which is generally intractable. This difficulty was addressed by exploiting an interesting property of the exponential family that relates the derivatives of the normalizing constant $C(\boldsymbol{\theta})$ to expectations of the MRF potential. Based on this, it was proposed to estimate the Fisher information matrix of the MRF (and therefore the CRB) using a Monte Carlo method. The proposed approach was successfully applied to the Ising and the Potts models, which are frequently used in signal processing applications. The resulting bounds have been used, in turn, to assess the statistical efficiency of three state-of-the art estimation methods that are interesting for image processing applications. An extension of the proposed method to hidden MRFs is currently under investigation. Perspectives for future work include the derivation of Bayesian Cramer Rao bounds for intractable models whose unknown parameters are assigned prior distributions.

\renewcommand{\baselinestretch}{1.20}
\footnotesize
\bibliographystyle{ieeetran}
\bibliography{strings}

\begin{thebibliography}{10}
\providecommand{\url}[1]{#1}
\csname url@samestyle\endcsname
\providecommand{\newblock}{\relax}
\providecommand{\bibinfo}[2]{#2}
\providecommand{\BIBentrySTDinterwordspacing}{\spaceskip=0pt\relax}
\providecommand{\BIBentryALTinterwordstretchfactor}{4}
\providecommand{\BIBentryALTinterwordspacing}{\spaceskip=\fontdimen2\font plus
\BIBentryALTinterwordstretchfactor\fontdimen3\font minus
  \fontdimen4\font\relax}
\providecommand{\BIBforeignlanguage}[2]{{%
\expandafter\ifx\csname l@#1\endcsname\relax
\typeout{** WARNING: IEEEtran.bst: No hyphenation pattern has been}%
\typeout{** loaded for the language `#1'. Using the pattern for}%
\typeout{** the default language instead.}%
\else
\language=\csname l@#1\endcsname
\fi
#2}}
\providecommand{\BIBdecl}{\relax}
\BIBdecl

\bibitem{Moller2006}
J.~Moller, A.~N. Pettitt, R.~Reeves, and K.~K. Berthelsen, ``{An efficient
  Markov chain Monte Carlo method for distributions with intractable
  normalising constants},'' \emph{Biometrika}, vol.~93, no.~2, pp. 451--458,
  June 2006.

\bibitem{Murray2006}
I.~Murray, Z.~Ghahramani, and D.~MacKay, ``{MCMC} for doubly-intractable
  distributions,'' in \emph{Proc. (UAI 06) 22nd Annual Conference on
  Uncertainty in Artificial Intelligence}, Cambridge, MA, USA, July 2006, pp.
  359--366.

\bibitem{Andrieu2009}
C.~Andrieu and G.~O. Roberts, ``The pseudo-marginal approach for efficient
  {M}onte {C}arlo computations,'' \emph{Ann. Statist.}, vol.~37, no.~2, pp.
  697--725, 2009.

\bibitem{Andrieu2010}
C.~Andrieu, A.~Doucet, and R.~Holenstein, ``Particle {M}arkov chain {M}onte
  {C}arlo methods,'' \emph{J. Roy. Stat. Soc. Ser. B}, vol.~72, no.~3, May
  2010.

\bibitem{Descombes1999}
X.~Descombes, R.~Morris, J.~Zerubia, and M.~Berthod, ``Estimation of {M}arkov
  random field prior parameters using {M}arkov chain {M}onte {C}arlo maximum
  likelihood,'' \emph{IEEE Trans. Image Process.}, vol.~8, no.~7, pp. 945--963,
  June 1999.

\bibitem{Forbes2007}
F.~Forbes and G.~Fort, ``Combining {M}onte {C}arlo and mean-field-like methods
  for inference in hidden {M}arkov random fields,'' \emph{IEEE Trans. Image
  Process.}, vol.~16, no.~3, pp. 824--837, March 2007.

\bibitem{McGrory2009}
C.~{McGrory}, D.~Titterington, R.~Reeves, and A.~Pettitt, ``Variational {B}ayes
  for estimating the parameters of a hidden {P}otts model,'' \emph{Statistics
  and Computing}, vol.~19, no.~3, pp. 329--340, Sept. 2009.

\bibitem{Giovannelli}
J.-F. Giovannelli, ``Ising field parameter estimation from incomplete and noisy
  data,'' in \emph{Proc. IEEE Int. Conf. Image Proc. (ICIP)}, Sept. 2011, pp.
  1853 --1856.

\bibitem{Pereyra_TIP_2012}
M.~Pereyra, N.~Dobigeon, H.~Batatia, and J.-Y. Tourneret, ``Estimating the
  granularity parameter of a {P}otts-{M}arkov random field within an {MCMC}
  algorithm,'' \emph{IEEE Trans. Image Process.}, vol.~22, no.~6, pp.
  2385--2397, June 2013.

\bibitem{Everitt2012}
R.~G. Everitt, ``Bayesian parameter estimation for latent {M}arkov random
  fields and social networks,'' \emph{J. Comput. Graphical Stat.}, 2012, to
  appear.

\bibitem{VanTreesBook}
H.~L.~V. Trees, \emph{Detection, estimation, and modulation theory: {P}art
  I}.\hskip 1em plus 0.5em minus 0.4em\relax New York: Wiley, 1968.

\bibitem{LiBook}
S.~Z. Li, \emph{{Markov random field modeling in image analysis}}.\hskip 1em
  plus 0.5em minus 0.4em\relax Secaucus, NJ, USA: Springer-Verlag New York,
  Inc., 2001.

\bibitem{Robert}
C.~P. Robert and G.~Casella, \emph{Monte Carlo Statistical Methods}.\hskip 1em
  plus 0.5em minus 0.4em\relax New York: Springer-Verlag, 1999.

\bibitem{BayesianChoice}
C.~P. Robert, \emph{The Bayesian Choice: From Decision-Theoretic Foundations to
  Computational Implementation (2nd ed.)}.\hskip 1em plus 0.5em minus
  0.4em\relax New York: Springer-Verlag, 2001.

\bibitem{Bishop}
C.~M. Bishop, \emph{Pattern recognition and machine learning}.\hskip 1em plus
  0.5em minus 0.4em\relax New York: Springer-Verlag, 2006.

\bibitem{MacKay}
D.~M. Kay, \emph{Information theory, Inference and learning Algorithms}.\hskip
  1em plus 0.5em minus 0.4em\relax Cambridge University Press, 2003.

\bibitem{Geyer92}
C.~J. Geyer, ``Practical {M}arkov chain {M}onte {C}arlo,'' \emph{Statistical
  Science}, vol.~7, no.~4, pp. 473--483, Nov. 1992.

\bibitem{Swendsen87}
R.~Swendsen and J.~Wang, ``Nonuniversal critical dynamics in {M}onte {C}arlo
  simulations,'' \emph{Physical Review Letters}, vol.~58, no.~2, pp. 86 -- 88,
  Jan. 1987.

\bibitem{Vincent2010}
T.~Vincent, L.~Risser, and P.~Ciuciu, ``Spatially adaptive mixture modeling for
  analysis of f{MRI} time series,'' \emph{IEEE Trans. Med. Imag.}, vol.~29,
  no.~4, pp. 1059 --1074, April 2010.

\bibitem{Eches2010tgrs}
O.~Eches, N.~Dobigeon, and J.-Y. Tourneret, ``Enhancing hyperspectral image
  unmixing with spatial correlations,'' \emph{IEEE Trans. Geoscience and Remote
  Sensing}, vol.~49, no.~11, pp. 4239--4247, Nov. 2011.

\bibitem{PereyraTMIC2011}
M.~Pereyra, N.~Dobigeon, H.~Batatia, and J.-Y. Tourneret, ``Segmentation of
  skin lesions in 2{D} and 3{D} ultrasound images using a spatially coherent
  generalized {R}ayleigh mixture model,'' \emph{IEEE Trans. Med. Imaging.},
  vol.~31, no.~8, pp. 1509--1520, Aug. 2012.

\bibitem{Pereyra_TIP_TechReport_2012}
\BIBentryALTinterwordspacing
------, ``Estimating the granularity parameter of a {P}otts-{M}arkov random
  field within an {MCMC} algorithm,'' University of Toulouse,
  IRIT/INP-ENSEEIHT, France, Tech. Rep., Feb. 2012. [Online]. Available:
  \url{http://pereyra.perso.enseeiht.fr/pdf/PereyraIEEETIPtr2012.pdf}
\BIBentrySTDinterwordspacing

\end{thebibliography}
\end{document}